\documentclass[pre,nofootinbib,floatfix,superscriptaddress, twocolumn]{revtex4}
\usepackage{dcolumn}
\usepackage{graphicx}
\usepackage{amsmath,amsfonts,amssymb, epsfig}
\usepackage{amsthm}
\usepackage{natbib}
\usepackage[outdir=.]{epstopdf}
\usepackage{verbatim}
\usepackage[caption=false, position=top]{subfig}
\usepackage{color}
\usepackage{rotating}
\usepackage{float}

\usepackage[pdftex,bookmarks, colorlinks, citecolor =blue, breaklinks]{hyperref}

\newcommand{\twopartdef}[8]
{
	\left\{
		\begin{array}{ll}
			#1 & \mbox{#2 } #3 \enspace#7 \\
			#4 & \mbox{#5 } #6 \enspace#8
		\end{array}
	\right.
}

\begin{document}
\title{Dynamic regulation of resource transport induces criticality in interdependent networks of excitable units}
\author{Yogesh S. Virkar}
\email{Yogesh.Virkar@colorado.edu}
\affiliation{University of Colorado at Boulder, Boulder, CO, 80309, USA}
\author{Juan G. Restrepo}
\email{Juan.Restrepo@colorado.edu}
\affiliation{University of Colorado at Boulder, Boulder, CO 80309, USA}
\author{Woodrow L. Shew}
\affiliation{University of Arkansas, Fayetteville, AR 72701, USA}
\author{Edward Ott}
\affiliation{University of Maryland, College Park, MD 20742, USA}

%%%%%%%%%%%%%%%%%%%%%%%%%%%%%%%%%%%

% 0. Abstract

% 1. Introduction

% 2. Model 

% 3. Experimental setup

% 4. Experiment 1: Self-organization to criticality. 

% 5. Experiment 2: Avalanche size distributions follow the power-law distribution. 

% 6. 3-dimensional map: both variants  

% 7. 3-dimensional map without noise: constraints

% 8. 3-dimensional map: Result using numerical experiment. 

% 9. Experiment 3: Heterogeneous glial source rates. 

% 10. Conclusions. 

%%%%%%%%%%%%%%%%%%%%%%%%%%%%%%%%%%%

%%%%%%%%%%%%%%%%%%%%%%%%%
%                           ABSTRACT                            %
%%%%%%%%%%%%%%%%%%%%%%%%%
\begin{abstract}

Various functions of a network of excitable units can be enhanced if the network is in the `critical regime', where excitations are, on average, neither damped nor amplified. An important question is how can such networks self-organize to operate in the critical regime. Previously it was shown that regulation via resource transport on a secondary network can robustly maintain the primary network dynamics in a balanced state where activity doesn't grow or decay. Here we show that this inter-network regulation process robustly produces a power-law distribution of activity avalanches, as observed in experiments, over ranges of model parameters spanning orders of magnitude. We also show that the resource transport over the secondary network protects the system against the destabilizing effect of local variations in parameters and heterogeneity in network structure. For homogeneous networks, we derive a reduced 3-dimensional map which reproduces the behavior of the full system.

%More specifically, we consider the example of a neural network with neurons (nodes) and synapses (edges). A  derive a model where synapse strengths are regulated by metabolic resources distributed by a secondary network of glial cells. We find that this two-layer network robustly preserves the critical state and produces power-law distributed avalanches over a wide range of parameters. In addition, the glial cell network protects the system against the destabilizing effect of local variations in parameters and heterogeneity in network structure. For homogeneous networks, we derive a reduced 3-dimensional map which reproduces the behavior of the full system.

\end{abstract}

\maketitle

%%%%%%%%%%%%%%%%%%%%%%%%%
%                        INTRODUCTION                      %
%%%%%%%%%%%%%%%%%%%%%%%%%
\section{Introduction}

Networks of excitable units are found in varied disciplines such as social science \cite{pei:tang:zheng:2015}, neuroscience \cite{larremore:shew:restrepo:2011}, epidemiology \cite{newmann:2003}, genetics \cite{newmann:2003}, etc. Various aspects of network function can be optimized when the network operates in the `critical regime', between low and high firing rates, as in neural networks \cite{shew:plenz:2013}, or at the `edge of chaos', between order and disorder, as in gene networks \cite{kauffman:1969}. In particular, for neural networks, criticality results in potential information handling benefits \cite{shew:plenz:2013}. 
A natural question receiving much interest \cite{levina2007dynamical,brochini2016phase,virkar:et:al:2016, kinouchi2019stochastic} is what mechanisms can lead such complex and distributed systems to operate in the critical regime, which typically occurs in a very small region of parameter space. In Ref. \cite{virkar:et:al:2016} we proposed a general mechanism based on the regulation of the excitable network dynamics by a resource which enables the interactions between the excitable elements and that is transported across a secondary network.  However, it was not clear if resource transport regulation is enough to produce experimental signatures of critical dynamics such as power-law distributions of avalanche sizes. In addition, the robustness of the model to parameter choices was not understood. Here we show that resource transport regulation leads to power-law distributed avalanche size distributions over model parameter ranges spanning orders of magnitude, and we validate our observations with a theoretical analysis which could serve as a basis to study more refined models of resource transport regulation. As a concrete case, we focus on the case of neural networks, where metabolic resources that facilitate the transmission of neural excitations are transported across a secondary glial network \cite{Giaume:2010fj, Fields:2015yq,Suzuki:2011rt, giaume:et:al:2010}. We emphasize, however, that our results could be applicable to other systems that operate at or near a critical point \cite{kauffman:1969, kitzbichler2009broadband, krishnamachari2001phase}.

%In this Letter, we derive a general mechanism based on the regulation of the excitable network dynamics by a resource which enables the interactions between the excitable elements and that is transported across a secondary network. As a concrete case, we will focus on the case of neural networks, where metabolic resources that facilitate the transmission of neural excitations are transported across a secondary glial network \cite{Giaume:2010fj, Fields:2015yq,Suzuki:2011rt, giaume:et:al:2010}. We find that resource-transport dynamics induces critical neural network dynamics characterized by neuronal avalanches following a power-law distribution for a wide range of model parameters. Finally, we show that in the presence of heterogeneity in local parameters or network structure, resource-transport amongst the glial cells can be essential to maintain criticality.

%In this Letter, we propose a model consisting of two layers: an excitatory neural network coupled to a complementary glial network that provides metabolic resources required for neuron firing. We find that resource-transport dynamics induces critical neural network dynamics characterized by neuronal avalanches following a power-law distribution. This critical state is shown to be robust to a wide range of parameter changes. Finally, in the presence of heterogeneity in local parameters or network structure, resource-transport amongst the glial cells is a requisite for maintaining criticality.  

%%%%%%%%%%%%%%%%%%%%%%%%%
%                               MODEL                              %
%%%%%%%%%%%%%%%%%%%%%%%%%
\section{Model}

\begin{figure} 
\centering{
\subfloat{{\label{fig:model}}\includegraphics[scale=0.35]{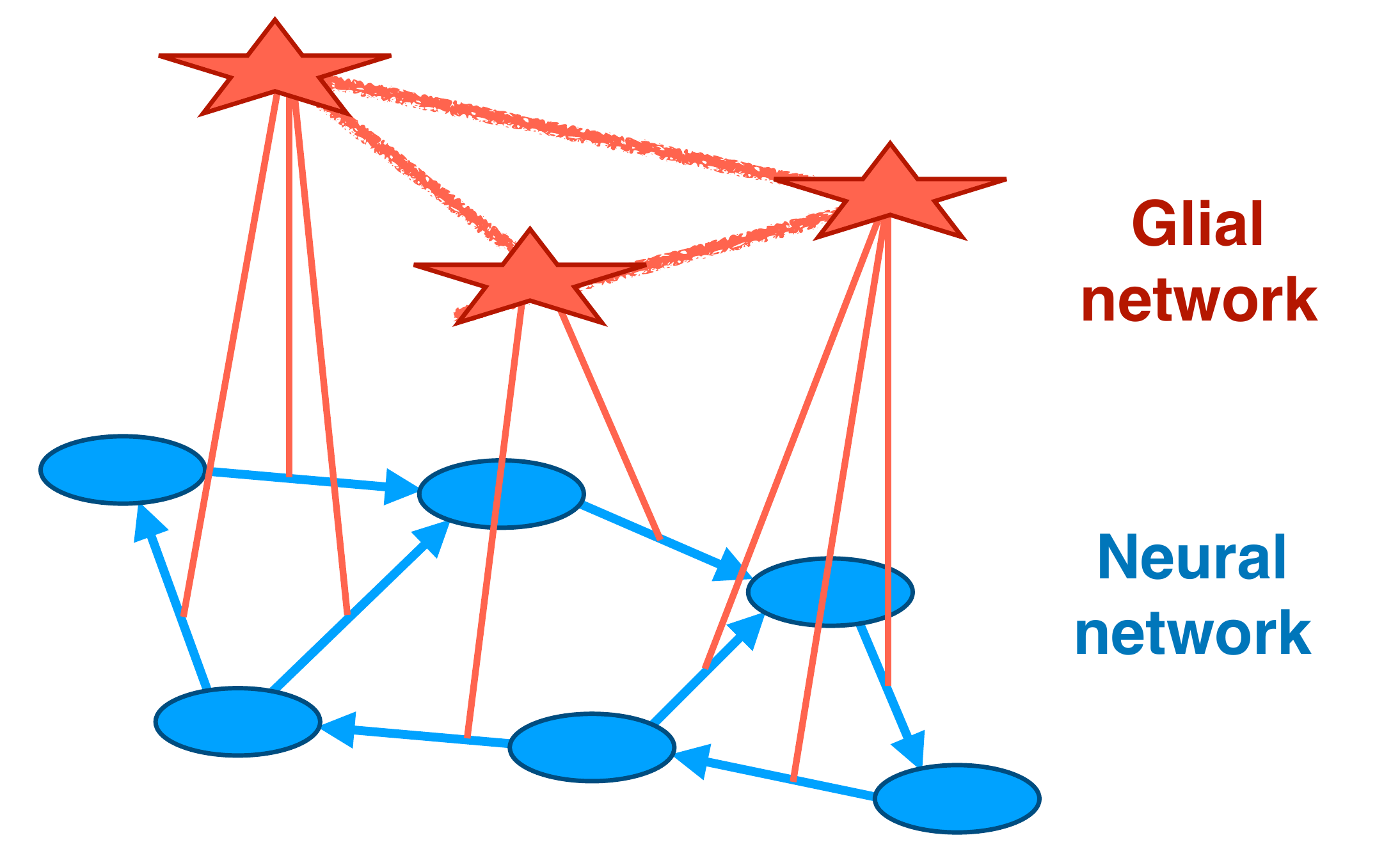}}\\
}
\caption{Our model consists of interacting neural and glial networks. The neural network is directed and weighted while the glial network is undirected and unweighted. }
\label{fig:model_specs}
\end{figure}

Following \cite{virkar:et:al:2016}, we consider a network model with two interdependent networks: a weighted directed neural network, and an unweighted undirected glial network which transports and regulates the supply of resources needed for the functioning of the neural network (see Fig.~\ref{fig:model_specs}). 
%(see Fig.~\ref{fig:model_specs}). 

{\it Neural network dynamics}: The neural network consists of $N$ excitable nodes that represent neurons, labeled $n = 1,2,\dots, N$, and $M$ directed edges (each corresponding to a synapse) and labeled $\eta = 1,2, \dots, M$. We also indicate a synapse pointing from node $m$ to node $n$ by $\eta_{(n,m)}$. At each discrete time step, $t$, neuron $n$ is in either the quiescent state ($s_{n}^t = 0$) or the active state ($s_{n}^t = 1$). We define $W^t$ as the $N \times N$ adjacency matrix whose entry $W_{nm}^t$ denotes the weight of the synapse on the edge from neuron $m$ to neuron $n$ at time $t$. 
The state of neuron $n$, $s_n^t$, is updated probabilistically based on the sum of its synaptic input from active presynaptic neurons in the previous time step, 
\begin{align} \label{eq:sn_t_1}
s_n^{t+1} = \twopartdef{1}{with probability}{\sigma \left(\displaystyle\sum_{m=1}^{N} W_{nm}^t s_m^t + \mu\right)} {0} {otherwise} {}{,}{.}
\end{align}
As in Refs.~\cite{larremore:et:al:2014,virkar:et:al:2016}, the model transfer function probability $\sigma$ is piecewise linear; $\sigma(x) = 0$ for $x \leq 0$, $\sigma(x) = x$ for $0 < x < 1$, and $\sigma(x) = 1$ for $x \geq 1$, and $\mu = 1/15000$ is a small external input that allows the system to avoid getting trapped in the absorbing state $s_n = 0$ for all $n$.

At time $t$ each synapse $\eta$ is assumed to have a supply $R_{\eta}^t$ of a metabolic resource, some of which is consumed every time the presynaptic neuron, $m(\eta)$, fires. While in this paper we do not focus on a particular resource, we note that $R$ could represent various metabolites that are transported diffusively among the glial cells, such as glucose and lactate \cite{rouach:et:al:2008}. Reflecting the increasing synaptic firing ability with increasing resource, we assume that the weight $W_{nm}^t$ in synapse $\eta_{(n, m)}$ from neuron $m$ to neuron $n$ is proportional to the amount of resource in the synapse, $W_{nm}^t = w_{nm} R_{\eta(n,m)}^t$ \cite{virkar:et:al:2016}. Finally, for simplicity, we consider only excitatory neurons and assume that there is no learning (these were considered in \cite{virkar:et:al:2016}). Thus, synaptic weight changes are caused only by the dynamics of resource transport.

The second network of our model, the unweighted and undirected glial network, consists of $T$ glial cells labeled $i = 1,2,\dots, T$. Each glial cell $i$ holds an amount of resource $R_{i}^t$ at time step $t$. Resources diffuse between the glial cells that are connected to each other. We define a $T \times T$ symmetric glial adjacency matrix $U$ such that entry $U_{ij} = 1$ if glial cell $j$ is connected to glial cell $i$ and $U_{ij} = 0$ otherwise. Each glial cell serves a set of synapses by supplying resource to them. Hence we define a $T \times M$ matrix $G$ with entries $G_{i\eta} = 1$ if glial cell $i$ serves synapse $\eta$ and $G_{i \eta} = 0$ otherwise. 
%Consistent with recent experimental studies \cite{halassa:et:al:2007}, we assume that all the incoming synapses of each neuron (i.e., its dendrites) are served by a single glial cell and that this glial cell serves no other neurons. So, given a synapse $\eta$, there is a unique glial cell $i(\eta)$ such that $G_{i(\eta)\eta}$ = $1$. 

{\textit{Resource-transport dynamics}}: Resource diffuses between glia through their connection network (characterized by the adjacency matrix $U$) and between glia and the synapses they serve (via the glial-neural connection network characterized by the adjacency matrix $G$). Our model for the evolution of the amount of resource $R_i^t$ at glial cell $i$ and the amount of resource $R_{\eta}^t$ at synapse $\eta$ is 
\begin{align}\label{eq:diffusion_betweenGlia}
R_i^{t+1} &= R_i^t + C_1 + D_G \displaystyle\sum_{j=1}^T U_{ij} \left(R_j^t - R_i^t\right) \nonumber \\ &+ D_S \displaystyle\sum_{\eta = 1}^{M} G_{i\eta} \left(R_{\eta}^t - R_{i}^t \right) \enspace,
\end{align}
\begin{align}\label{eq:diffusion_betweenGliaSynapse}
R_{\eta}^{t+1} = R_{\eta}^t + D_S \left(R_{i(\eta)}^t - R_{\eta}^t \right) - C_2 s_{m(\eta)}^t \enspace,
\end{align}
where $D_G$ is the rate of diffusion between glial cells, and $D_S$ is the rate of diffusion between glia and synapses. Moreover, we enforce $R_{\eta} \geq 0$, i.e., if Eq.~\eqref{eq:diffusion_betweenGliaSynapse} yields $R_{\eta}^{t+1} < 0$, then we replace it by $0$. The model parameter $C_1$ on the right hand side of Eq.~\eqref{eq:diffusion_betweenGlia} denotes the amount of resource added to each glial cell at each time step  (e.g., supplied by capillary blood vessels). For simplicity, we assume first that each glial cell has the same $C_1$ (the effect of heterogeneous values of $C_1$ will be discussed later). The last two terms in Eq.~\eqref{eq:diffusion_betweenGlia} are the amount of resource transported to glial cell $i$, respectively, from its neighboring glial cells and from the synapses that it serves.

%The first term on the right hand side of \eqref{eq:diffusion_betweenGliaSynapse} denotes the amount of resource at synapse $\eta$ at time $t$. 
The term proportional to $D_S$ in Eq.~\eqref{eq:diffusion_betweenGliaSynapse} denotes the amount of resource gained (if $R_{i(\eta)}^t > R_{\eta}^t$) or lost (if $R_{i(\eta)}^t < R_{\eta}^t$) from glial cell $i(\eta)$ that serves synapse $\eta$. If the presynaptic neuron $m(\eta)$ fires at time step $t$ ($s_{m(\eta)}^t = 1$), then synapse $\eta$ consumes an amount of resource $C_2$, thus decreasing the resource at synapse $\eta$ by this amount.

%%%%%%%%%%%%%%%%%%%%%%%%%%%%%%%%%%%%%%%%%%%%
% TODO: Instead of setting, use C_1 on the "x"-axis. This result needs to tie in to the 
%             result from the 3-dimensional map (Fig.4). The same parameter settings should 
%             be used. 
%%%%%%%%%%%%%%%%%%%%%%%%%%%%%%%%%%%%%%%%%%%%

%%%%%%%%%%%%%%%%%%%%%%%%%
%                EXPERIMENTAL SETUP                 %
%%%%%%%%%%%%%%%%%%%%%%%%%
\section{Numerical Experiments}

We now describe and present the results of numerical experiments on our model, Eqs.~\eqref{eq:sn_t_1}-\eqref{eq:diffusion_betweenGliaSynapse}. Our main goal is to show that resource transport dynamics robustly regulates the operation of the neural network in the critical regime. In the neural model used here, the critical regime is characterized by the largest eigenvalue of the neural synapse matrix $W^t$, $\lambda^t$, being one \cite{larremore:et:al:2014,virkar:et:al:2016}. Therefore we will consider  $\lambda^t \approx 1$ as one criterion for criticality. However, a more practical definition of criticality, applicable more generally \cite{shew:plenz:2013}, is a power-law distribution of the sizes of activity bursts, or {\it neuronal avalanches}. We will also verify that the model {\it robustly} produces power-law distributed neuronal avalanches.

In our numerical experiments, we consider an Erd\"os-R\'enyi network structure for both the neural and glial networks.  The neural network is described using an $N \times N$ adjacency matrix, $W$, such that with probability $p$ we have an entry $w_{nm} \neq 0$ that represents a link from node $m$ to node $n$.
%and otherwise we have an entry $w_{nm} = 0$ that represents the absence of such a link. 
At time $t=0$, we set the resource at each synapse $\eta$, $R_{\eta}^0 = 1$, and draw $w_{nm}$ from  a uniform distribution over $[0,\bar w]$. By choosing the value of $\bar w$, we can set the initial largest eigenvalue of $W^t$, i.e., $\lambda^0$, to a desired value, and test whether the subsequent evolution of the model results in $\lambda^t \to 1$.

%\begin{table}[b]
%\begin{tabular} [b]{ c c c c c c }
%\hline \\
%Layer & Nodes & Adjacency & Prob. of & Mean  & Mean no.  \\
% &  &matrix & an edge & degree & of edges \\
%\hline \\
%Neural & $N$ & $W$ & $p$ &  $k_N = N p $ & $M = N k_N$ \\
%&1000&&0.05&50&50000\\ 
%Glial  &$T$ & $U$ & $q$ & $k_G = T q$ & $E = T k_G / 2$ \\ 
%&1000&&0.05&50&25000\\ 
%\hline 
%\end{tabular}
%\caption{Parameters for the neural and glial networks.}
%\label{tab:neural_glial}
%\end{table}
The glial network is undirected and unweighted and its adjacency matrix is given by a $T \times T$ matrix, $U$, such that with probability $q$ we have an entry $U_{ij} = U_{ji} = 1$ that represents an undirected link between nodes $i$ and $j$. Motivated by experiments \cite{azevedo:et:al:2009} showing $T \sim N$, for specificity, we take the number of glia and neurons to be equal, $T=N$. Consistent with this, and the additional experimental finding that all incoming synapses of a given neuron are served by the same glial cell \cite{halassa:et:al:2007}, we further assume that each glial cell serves a unique neuron. We set the initial resource for each glial cell to be equal to an uniform value, $R_i^0 = r$ (note that $R_{\eta}^0$ may be different from $R_i^0$).  For all numerical experiments we use the values $T = N = 1000$, $p = q = 0.05$, and assume, for simplicity, that the entries of matrices $U$ and $W$ are independent of each other. Unless mentioned otherwise, the parameters for resource-transport dynamics are set as $D_G = D_S = 5 \times 10^{-5}$,  $C_1 = 6 \times 10^{-8}$, and $C_2 = 10^{-8}$. 
%Starting with different initial conditions (e.g., $\lambda^0 = 1, 0.98,1.02$), the resource-transport dynamics causes the system to self-organize to $\lambda^t = 1$ after a transient period. Although we chose the parameter values somewhat arbitrarily, we will show that our findings are fairly robust to these values.
%, and that we can determine the parameter ranges for which our model results in $\lambda^t \approx 1$ from a simplified 3-dimensional map. 

%%  Mention the particular values for N, p, T, q, etc. 
%% Parameter choices generally used. 

%%%%%%%%%%%%%%%%%%%%%%%%%%%%
%                 EXPERIMENT 1: Explanation
%%%%%%%%%%%%%%%%%%%%%%%%%%%%

\begin{figure} [t]
\centering{
\subfloat{{\label{fig:stability_critical_lambda}}\includegraphics[scale=0.36]{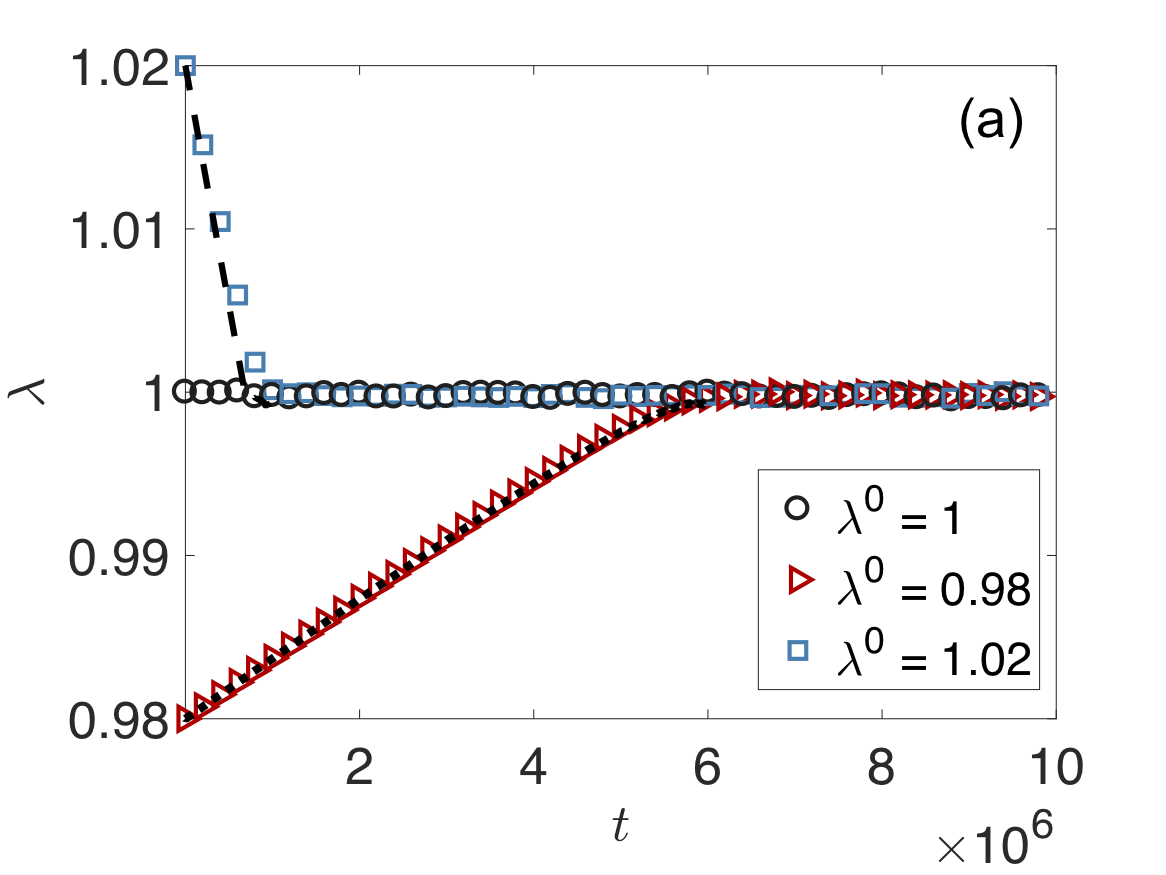}} \\
\subfloat{{\label{fig:stability_critical_avgRi}}\includegraphics[scale=0.38]{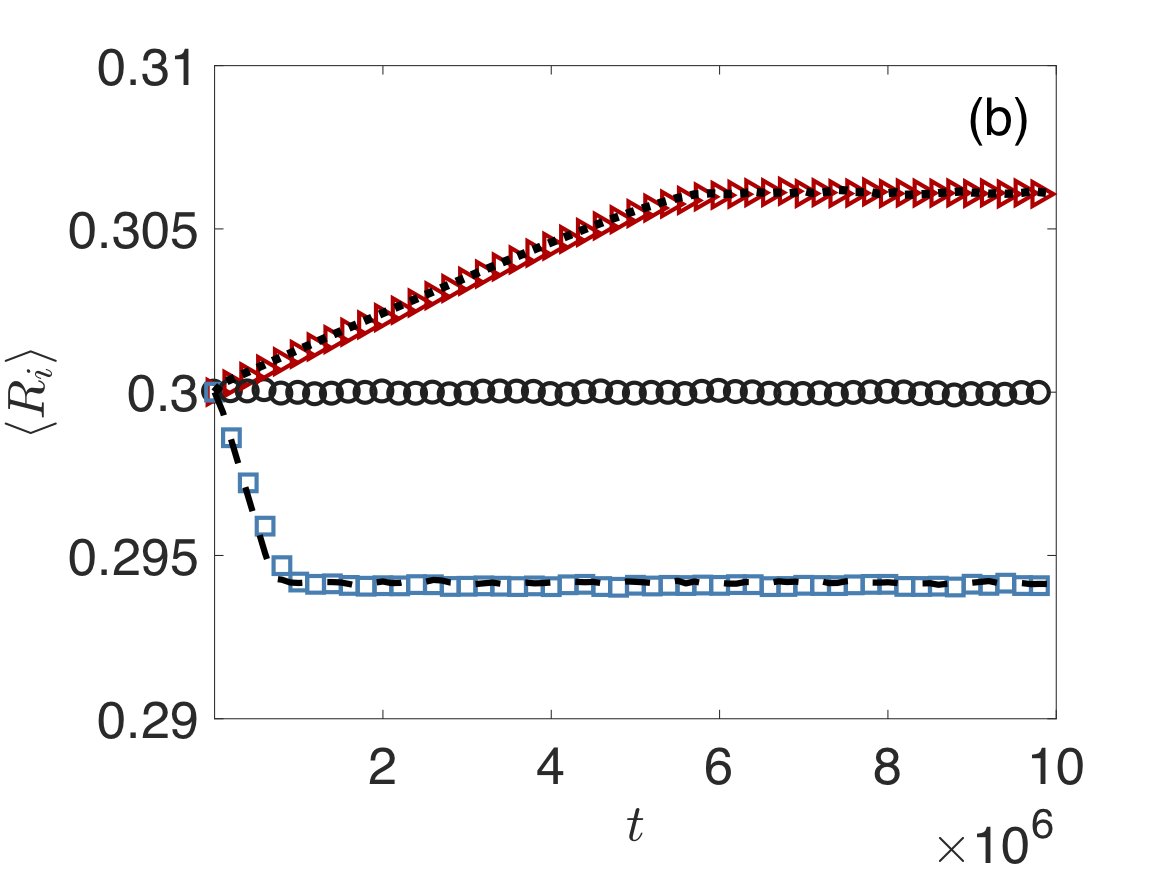}}
}
\caption{(a) Largest eigenvalue, $\lambda^t$, of the neural network adjacency matrix for three initial conditions $\lambda^0 = 1, 0.98,1.02$ (black circles, red triangles, and blue squares, respectively) as a function of time. 
(b) Average glial resource $R^t = \sum_i R^t_i/N$ as a function of time for the same initial conditions. In both panels, the dotted (subcritical case) and the dashed (supercritical case) lines show the predictions from the \textit{3-D map with noise} [Eqs.~(\ref{eq:3DmapR}), (\ref{eq:3Dmaplambda}),(\ref{eq:3DmapS_noise})].
}
\label{fig:model_results}
\end{figure}

In the first experiment, we show that starting with different initial conditions $\lambda^0 = 1, 0.98,1.02$, the resource-transport dynamics causes the system to self-organize to the critical state corresponding $\lambda^t = 1$ after a transient period. In Fig.~\ref{fig:model_results} (a) we show $\lambda^t$ for the three different initial conditions $\lambda^0 = 1, 0.98,1.02$ (black circles, red triangles, and blue squares, respectively). In the three cases, $\lambda^t$ approaches and subsequently remains close to $1$ (this will be quantified in Fig.~\ref{fig:oscillations_deltalambda}).
In Fig.~\ref{fig:model_results} (b) we show that the average glial resource, $R^t = \sum_i R^t_i/N$, reaches a steady state in all three cases.

%%%%%%%%%%%%%%%%%%%%%%%%%%%%
%                 EXPERIMENT 2: Explanation
%%%%%%%%%%%%%%%%%%%%%%%%%%%%

% Experimental setup: 
% 1. Explain parameter settings used. Their relation to the constraints. 
% 2. What we plot on z-axis, i.e., P(L). 

% Methodology. 

% 1. Computing the avalanche sizes. 
% 2. Computing Fitting using maximum likelihood methods \cite{clauset:shalizi:newman:2009}
% 3. Goodness-of-fit using the KS-statistic and thresholding per the rule used in \cite{woodrow:et:al:2015} turtles paper. 
 
% Result:
% 1. As expected robust power laws are observed over a large range of input for all settings. 
% 2. However, despite what the constraints predict, the observed power-law behavior holds for smaller ranges of avalanche sizes as we increase C_1, while keeping C_2 fixed. 
% 3. This probably happens because as C_1 increases the noise in \lambda increases. 
% 4. Connect this to the experiment involving 3-dimensional map. 

As discussed above, we are interested in whether the dynamics of the neuronal network reproduces experimental signatures of critical behavior, in particular power-law distributed avalanches of activity.
%Though in the above experiment, we characterize the critical state regime by $\lambda \approx 1$, in other works (for e.g. \cite{beggs:plenz:2003, shew:plenz:2013, larremore:shew:restrepo:2011}), it is often characterized by avalanches with sizes following a power-law distribution with the characteristic exponent of $-3/2$. Since the inclusion of inhibition causes the dynamics of $S$ to be ceaseless \cite{larremore:et:al:2014}, 
To do this, following \cite{larremore:et:al:2014}, we define a measure of activity, $S(t) = \sum_{m} s_m^t/N$, and define an avalanche as the excursion of activity $S^t$ above a threshold $S^{*}$, i.e., $S^t < S^*$ for $t < t_1$, $t>t_2$ and $S^t \geq S^*$ for $t_1 \leq t \leq t_2$). We define the size $L$ of the avalanche as $L = N \displaystyle\sum_{t=t_1}^{t_2} S^t$, the number of spikes (excitations) over a single excursion.

\begin{figure}[b]
\centering{
\includegraphics[scale=0.38]{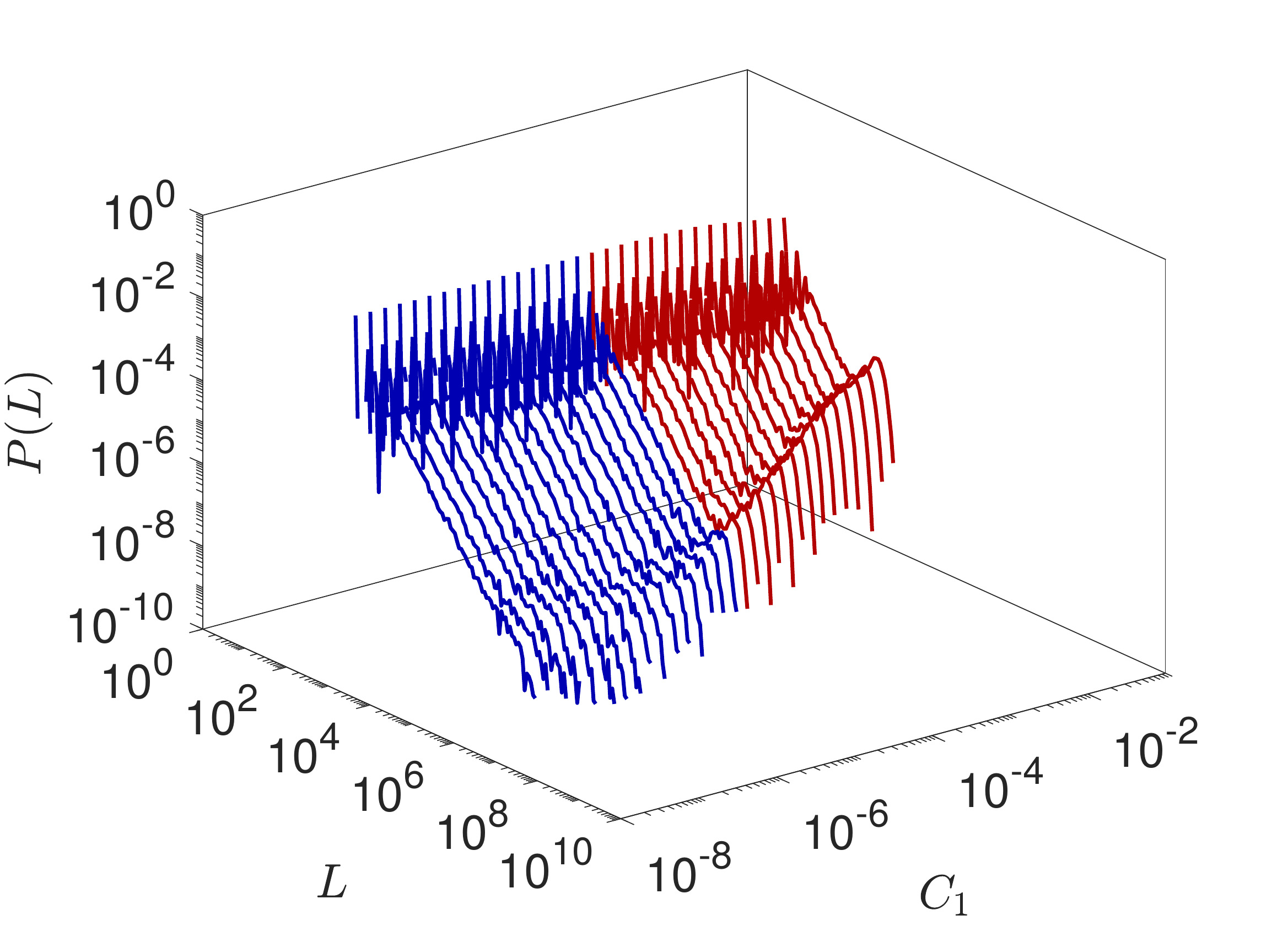}
}
%\caption{The critical state is robust to changes in model parameters. For a fixed $C_2=10^{-8}$, panel (a) shows the avalanche size distributions, $P(L)$ for $9$ different parameter settings formed from the combinations of the following parameter choices: $C_1 = \left[3 \times 10^{-8}, 6 \times 10^{-8}, 12 \times 10^{-8}\right]$ and $D = \left[2.5 \times 10^{-5}, 5 \times 10^{-5}, 10^{-4}\right]$. As shown, the exponent for the power-law fit ($\gamma$) is near the characteristic value of $-3/2$ for fits spanning $3$ (red) or $4$ (blue) decades.}
\caption{Size distributions $P(L)$ for various values of $C_1$. Blue curves indicate plausible power-law fits (under $10\%$ level of significance) with $P(L) \propto L^{\gamma}$ such that $\gamma \approx -3/2$ and the red curves indicate rejected  power-law fits. }
\label{fig:robustness_critical_state}
\end{figure}

%Fig.~\ref{fig:robustness_critical_state} shows that for $S^{*} = 0.15$, the critical state regime as characterized by avalanche size distributions $P(L) \propto L^{-3/2}$ is robustly observable over a wide range of parameter settings. The caption of Fig.~\ref{fig:robustness_critical_state} shows the parameter settings used and the corresponding power-law fit exponent, $\gamma$. Given a choice of $C_2$, we find that choosing a $C_1$ such that $C_1 < C_2 k_I$, where $k_I$ denotes the average number of synapses served by a glial cell, results in stable critical state. If $C_1$ does not satisfy this inequality, the system receives more resource than what the synapses can burn, resulting in the supercritical state. Empirically, we find that, for the tested range $C_1=\left[3 \times 10^{-8}, 12 \times 10^{-8}\right]$ and $D =\left[2.5 \times 10^{-5}, 10^{-4}\right]$, choosing smaller values of $C_1$ results in power-law fits over more decades. Also, for the particular choices for $C_1$ and $C_2$, the parameter $D$ does not have any impact on criticality. Finally, choosing smaller values of $C_1$ results in power-law fits spanning $4$ decades (shown in blue in Fig.~\ref{fig:robustness_critical_state}), while larger $C_1$ reduces the fitting range to $3$ decades (shown in red in Fig.~\ref{fig:robustness_critical_state}). Finally power-law fitting was performed using standard techniques \cite{clauset:shalizi:newman:2009} based on maximum-likelihood estimation (MLE) and hypothesis test using the Kolmogorov-Smirnov (KS) statistic. 

To investigate the robustness of our model to changes in parameters, we fix $D= D_G = D_S = 5 \times 10^{-5}$ and vary $C_1$ and $C_2$ logarithmically roughly from $10^{-8}$ to $10^{-2}$ keeping the ratio $C_2/C_1 = 1/6$ constant. Using the threshold $S^{*} = 0.15$, we calculate avalanche size distributions, $P(L)$, for each parameter setting. We then fit a power-law model using standard techniques \cite{clauset:shalizi:newman:2009} based on maximum-likelihood estimation (MLE) and a hypothesis test that generates a $p$-value using the Kolmogorov-Smirnov (KS) statistic. Since we have finite-size effects, in addition to the lower size $L_{\text{min}}$ cutoff used in \cite{clauset:shalizi:newman:2009}, we introduce an upper cutoff $L_{\text{max}}$, i.e., we test the plausibility of a power-law model where we condition on avalanche sizes in a range $[L_{\text{min}},L_{\text{max}}]$ \cite{Langlois:2014ij, Shew:2015pb}. We accept as plausible power laws only those distributions for which this range spans at least three decades.

In Fig.~\ref{fig:robustness_critical_state} we show size distributions $P(L)$ for various values of $C_1$. The blue curves indicate a plausible power-law fit (under $10\%$ level of significance) with $P(L) \propto L^{\gamma}$ such that $\gamma \approx -3/2$ (exponents range from $-1.42$ to $-1.49$), and the red curves a rejected power-law fit. We find that there is a large range of values of the parameters (small values of $C_1$ and $C_2$ for which the system results in power-law distributed avalanches, but that for some parameter choices (larger values of $C_1$ and $C_2$) the distribution of avalanche sizes no longer satisfies our Kolmogorov-Smirnov power-law test.

%Fig.~\ref{fig:robustness_critical_state} shows that for $S^{*} = 0.15$, the critical state regime as characterized by avalanche size distributions $P(L) \propto L^{-3/2}$ is robustly observable over a wide range of parameter settings. We perform power-law fitting using standard techniques \cite{clauset:shalizi:newman:2009} based on maximum-likelihood estimation (MLE) and hypothesis test that generates a $p$-value using the Kolmogorov-Smirnov (KS) statistic. Fixing $D = 5 \times 10^{-5}$, we vary $C_1$ and $C_2$ proportionally. The parameter settings that satisfy the inequalities \eqref{eq:ineq1}-\eqref{eq:ineq5} (shown in blue), derived from the 3-dimensional map without noise, result in stable critical state. For these settings, we obtain a $p$-value greater than the chosen significance level $0.1$. For parameter settings not satisfying the inequalities, we get $p < 0.1$ and hence we reject the power-law hypothesis. Finally, we find that, for the tested range $C_1=\left[\right]$ and $C_2 =\left[\right]$, choosing smaller values of $C_1$ and $C_2$ results in power-law fits over more decades. 

%For example, as shown in panel (b), using smaller $C_1$ resulted in power-law distributions spanning $4$ (blue) decades as opposed to $3$ (red) decades.

%%%%%%%%%%%%%%%%%%%%%%%%%
%                   3-DIMENSIONAL MAP                  %
%%%%%%%%%%%%%%%%%%%%%%%%%
\section{3-Dimensional Map}

%In order  to gain an understanding of the mechanisms that lead to the critical regime and to determine conditions on the model parameters that result in critical behavior, we assume a homogeneous network structure for both the neural and the glial networks, i.e., the out-degree of each neuron in the neuronal network is $k$. In this case, we can approximate the largest eigenvalue $\lambda^t$ of the neural network adjacency matrix by the mean degree to obtain $\lambda^t \approx \frac{1}{N} \sum_{n,m} W_{nm}^t 
%= \frac{1}{N}\sum_{n,m} w_{nm} R^t_{\eta(n,m)} \approx \bar w \frac{1}{N}\sum_{\eta} R^t_{\eta}$, where we assumed that the weights $w_{nm}$ are uncorrelated with the resource $R_{\eta(n,m)}$ and defined $R^t = \frac{1}{T} \sum_{i} R_i^t$ (recall we are taking $N = T$). Summing Eq.~\eqref{eq:diffusion_betweenGlia} over $i$ and Eq.~\eqref{eq:diffusion_betweenGliaSynapse} over $\eta$, we obtain (see Supplemental Material for details)

In order  to gain an understanding of the mechanisms that lead to the critical regime and to determine conditions on the model parameters that result in critical behavior, we make the following assumptions: (i) the neural network is large, uncorrelated and homogeneous so that the Perron-Frobenius eigenvalue of the matrix with entries $W_{nm} = w_{nm} R_{\eta(n,m)}$ is well approximated by its average row sum, (ii) the intrinsic synapse weights $w_{nm}$ are uncorrelated with $R_{nm}$ or have a narrow distribution around their average $\langle w \rangle$ so that $\sum_{n,m} w_{nm} R_{\eta(n,m)}^t \approx \langle w \rangle \sum_{n,m} R_{\eta(n,m)}^t$, and (iii) the glial cells all serve the same number of synapses $k$ (or the distribution of the number of synapses served is narrow). While some of these assumptions could be relaxed and the theory generalized, we leave these considerations for future work. 

%%%%%%%%%%%%%%%%%%%%%
%%%%%%%%%%%%%%%%%%%%%
% Derivation from Supplemental section. 
%%%%%%%%%%%%%%%%%%%%%

First, we define the average amount of resource per glial cell at time $t$:
\begin{align}
R^t = \frac{1}{T}\sum_{i} R_i^t.
\end{align}
Averaging Eq.~\eqref{eq:diffusion_betweenGlia} over glial cells, we obtain
\begin{align}
R^{t+1} = R^t + C_1 +\frac{ D_S }{T} \sum_{\eta=1}^M R_{\eta}^t \sum_{i=1}^TG_{i\eta}  -  \frac{ D_S }{T} \sum_{i=1}^T R_{i}^t \sum_{\eta=1}^M G_{i\eta}.
\end{align}
From the assumption that each glial cell serves $k$ synapses, we have $\sum_{\eta=1}^M G_{i\eta} = k$. Furthermore, since each synapse is served by a unique glial cell, $\sum_{i=1}^TG_{i\eta} = 1$, and we obtain
\begin{align}
R^{t+1} = R^t + C_1 +\frac{ D_S }{T} \sum_{\eta=1}^M R_{\eta}^t -  k D_S R^t.\label{eq:s3}
\end{align}
The term $\sum_{\eta=1}^M R_{\eta}$ is the total amount of resource in the synapses. From the assumption that the fixed synapse weights $w_{nm}$ are uncorrelated with $R_{nm}$ (or that their distribution is sufficiently narrow), the total resource in the synapses can be related to the sum of all entries $W_{nm}^t = w_{nm} R_{nm}^t$ of the synapse weight matrix
\begin{align}
\sum_{\eta=1}^M R_{\eta}^t = \frac{1}{\langle w \rangle}\sum_{n,m} W_{nm}^t.
\end{align}
For large homogeneous, uncorrelated networks, the average row sum is an excellent approximation to the Perron-Frobenius eigenvalue, and so
\begin{align}
\sum_{\eta=1}^M R_{\eta}^t \approx \frac{N}{\langle w \rangle} \lambda^t,
\end{align}
and Eq.~(\ref{eq:s3}) becomes, using $N = T$ as discussed before,
\begin{align}
R^{t+1} = R^t + C_1 +\frac{ D_S }{\langle w \rangle} \lambda^t -  k D_S R^t\label{eq:3DmapR}\enspace.
\end{align}
Summing Eq.~\eqref{eq:diffusion_betweenGliaSynapse} over $\eta$ and multiplying  by $\langle w \rangle/N$ we get
\begin{align}
\lambda^{t+1} = \lambda^t  + \frac{D_S \langle w \rangle}{N} \sum_{\eta = 1}^M R^t_{i(\eta)} -D_S \lambda^t - \frac{C_2 \langle w \rangle}{N} \sum_{\eta=1}^M s_{m(\eta)}.
\end{align}
Since there is a single glial cell serving each synapse, and each synapse serves $k$ glial cells,  $\sum_{\eta = 1}^M R^t_{i(\eta)} = \sum_{i = 1}^T k R^t_{i} = T k R^t = N k R^t$. In addition, since each glial cell serves all the $k$ synapses of a single neuron, $\sum_{\eta=1}^M s_{m(\eta)} = k \sum_{m=1} s_m^t = k N S^t$. So we obtain
\begin{align} 
\lambda^{t+1} = \lambda^t  + D_S \langle w \rangle k R^t -D_S \lambda^t - C_2 \langle w \rangle k S^t\label{eq:3Dmaplambda}\enspace.
\end{align}
%%%%%%%%%%%%%%%%%%%%
%%%%%%%%%%%%%%%%%%%%

%\begin{align} 
%R^{t+1} &= R^t + C_1 +  \frac{D_S\lambda^t}{\bar{w}} - k D_S R^t \label{eq:3DmapR}\enspace, \\
%\lambda^{t+1} &=\lambda^t + k D_S \bar{w} R^t - D_S\lambda^t - C_2 \bar{w} k S^t \label{eq:3Dmaplambda}\enspace,
%\end{align}
The mean field equations \eqref{eq:3DmapR}, \eqref{eq:3Dmaplambda} need to be closed with an equation for the evolution of the average activity, $S^t$, which is a stochastic variable determined by Eq.~(\ref{eq:sn_t_1}). In order to obtain a tractable map, we model $S^t$ in two different ways: in the first one, we neglect stochastic effects and use the  deterministic approximation:
\begin{align} \label{eq:3DmapS_no_noise}
S^{t+1} = \lambda^t S^t \enspace.
\end{align}
This approximation is based on the fact that, except for values of $S$ close to $0$ or $1$, the expectation of $S^{t+1}$ calculated from Eq.~(\ref{eq:sn_t_1}) is $\lambda^t S^t$. This approximation neglects the nonlinear effects that keep $S^t$ below $1$, and thus should be interpreted only as a guide to determine the fixed points and their stability in the limit of vanishing stochastic effects [i.e., when the expected number of terms in the sum in Eq.~(\ref{eq:sn_t_1}) is large]. We refer to Eqs.~\eqref{eq:3DmapR}, \eqref{eq:3Dmaplambda} and (\ref{eq:3DmapS_no_noise}) as the \textit{3-D map without noise}. %While this approximation neglects the effects of noise and is valid only close to the fixed point, it is useful since the stability properties of the fixed point $\lambda = 1$ underlie the robustness of the critical state to changes in model parameters. 
A more realistic model for $S^t$ includes a stochastic noise term and a mechanism to enforce $0 \leq S^t \leq 1$:
%\label{itm:assump2} In this case, we model $S^t$ with a noise term $r^t$ and small external stimulus probability $\mu^t$ as follows
\begin{align} \label{eq:3DmapS_noise}
S^{t+1} = \max \left( 0, \min \left(1, \lambda^t S^t + r^t + \mu^t \right)\right) \enspace.
\end{align}
Here, $r^t$ is a Gaussian noise term with zero mean and standard deviation $\sqrt{S^t(1-S^t)/N}$, as estimated in \cite{larremore:et:al:2014}, while $\mu^t$ represents an external stimulus taken to be $1/N$ with probability $\zeta$ and $0$ otherwise, introduced to prevent $S^t$ from decaying to zero. Effectively, this stimulus excites one neuron every time step with probability $\zeta$. We refer to Eqs.~\eqref{eq:3DmapR}, \eqref{eq:3Dmaplambda} and \eqref{eq:3DmapS_noise} as the \textit{3-D map with noise}. This variant of the map is useful to predict the evolution of the macroscopic variable $R^t$ of the full model. As an example, in Fig.~\ref{fig:model_results} we show with dotted and dashed lines the predictions of the evolution of $\lambda^t$ and the average glial resource $R^t$ obtained from the {\it 3-D map with noise}. The predictions agree very well with direct simulations of the full model.
% and, being a three dimensional map, are much more computationally efficient. 
%The resource dynamics, although not shown, also is well predicted by the {\it 3-D map with noise}.

%%%%%%%%%%%%%%%%%%%%%%%%%
%    3-DIMENSIONAL MAP: CONSTRAINTS     %
%%%%%%%%%%%%%%%%%%%%%%%%%

%While the {\it 3-D map with noise} allows us to reproduce the behavior of the full model, 
The simplicity of the \textit{3-D map without noise} allows us to derive parameter constraints that must be satisfied for a stable critical state. In particular, the system of Eqs. \eqref{eq:3DmapR}, \eqref{eq:3Dmaplambda}, \eqref{eq:3DmapS_no_noise} has the fixed point 
\begin{align}
\bar \lambda = 1, \hspace{1cm} \bar S = \frac{C_1}{k C_2}, \hspace{1cm} \bar R =  \frac{C_1}{k D} +  \frac{1}{k \langle w \rangle}. \label{fp1}
\end{align}
The critical state $\lambda = 1$ is a fixed point of the deterministic map. Its stability is determined by whether the eigenvalues of the Jacobian of the map \eqref{eq:3DmapR}, \eqref{eq:3Dmaplambda}, \eqref{eq:3DmapS_no_noise} evaluated at the fixed point (\ref{fp1}) 
%\begin{align}
%J = \left[ \begin{array}{ccc}
%1 - q D&\displaystyle\frac{D}{\bar{w}}&0\\
%qD\bar{w}&1-D&-k C_2 \bar{w}\\
%0&\displaystyle\frac{C_1}{k C_2}&1
%\end{array}
 %\right]
%\end{align}
are inside the complex unit circle. Applying the Routh-Hurwitz criterion, we find that the fixed point is stable (which we interpret as robustness of the critical state) if 
\begin{align}
&D k - \frac{2}{3} < 0, \\
&\frac{1}{k D}  -\frac{1+k}{C_1 k \langle w \rangle} - \displaystyle\frac{3}{4} < 0,\\%\label{eq:ineq1}, \enspace,\\
%\label{eq:ineq2}\enspace,\\
&\displaystyle\frac{C_1 k D \langle w \rangle}{8} - \displaystyle\frac{C_1 \langle w \rangle}{4} + \displaystyle\frac{D k}{2} + \displaystyle\frac{D}{2} - 1 < 0\label{eq:ineq3},\\
&C_1^2 D^2 k^2 \langle w \rangle - 2 C_1^2 D k \langle w \rangle + C_1^2 \langle w \rangle\nonumber\\
&+ C_1 D^2 k^2  + C_1 D^2 k  - C_1 D  < 0.\label{eq:ineq4}\end{align}
In addition, since $0\leq\bar S\leq 1$, we have the additional inequality $C_1/(k C_2) < 1$,
which represents the constraint that the amount of resource supplied to glial cells per time step can not exceed the amount that can be consumed at the synapses. 
\begin{figure}[b]
\centering{
\includegraphics[scale=0.35]{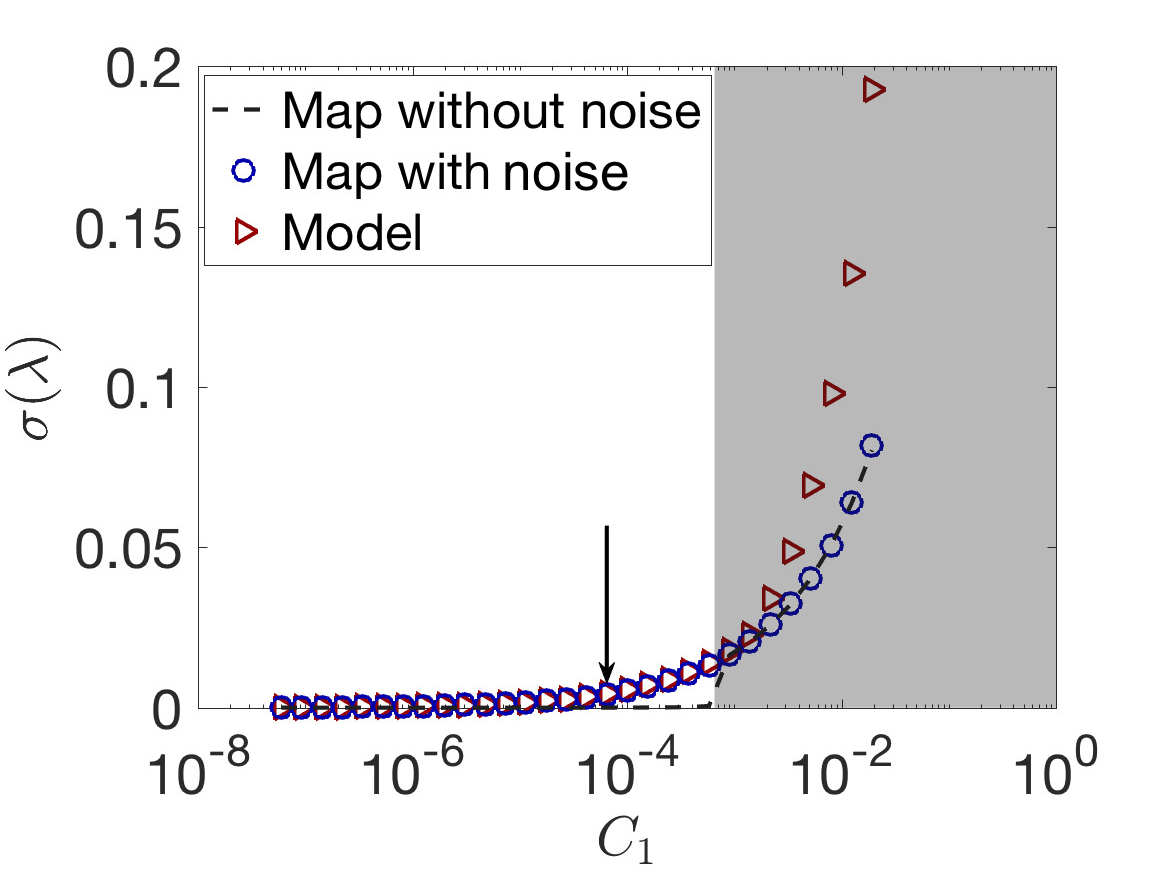}
}
\caption{Root mean squared deviation of $\lambda^t$ from $1$, $\sigma_\lambda = \sqrt{\langle (\lambda^t - 1)^2 \rangle_t}$ as a function of $C_1$. 
%(we take $C_2 = C_1/6$ and $D=5 \times 10^{-5}$).  The red triangles correspond to simulations of the full model, the blue circles to the {\it 3-D map with noise}, and the dashed line to the {\it 3-D map without noise}. 
The shaded grey region indicates parameter values for which $\bar \lambda = 1$ is linearly unstable. Values of $C_1$ to the left of the arrow yield avalanche size distributions that have plausible power-law fits with exponent close to $-3/2$.}
\label{fig:oscillations_deltalambda}
\end{figure}

To demonstrate the usefulness of these inequalities, we verify which of the curves in Figure~\ref{fig:robustness_critical_state} correspond to parameters which satisfy these inequalities. Parameters that satisfy (don't satisfy) the inequalities approximately correspond to distributions which follow (don't follow) a power-law. 
%We note, however, that in other cases, even though the fixed point $\lambda = 1$ is linearly unstable, $\lambda^t$ can oscillate so close about $1$ that the avalanche size distribution is still a plausible power law. Conversely, for moderately large networks finite size effects can lead to fluctuations of $\lambda^t$ about $1$ that are significant enough to cause deviations from power law behavior, as noted in Ref.~\cite{costa:et:al:2015}. 
To illustrate this, we plot in Fig.~\ref{fig:oscillations_deltalambda} the quantity $\sigma_\lambda = \sqrt{\langle (\lambda^t - 1)^2 \rangle_t}$, which measures the deviation of the system from $\lambda = 1$, as a function of $C_1$ (we take $C_2 = C_1/6$, $D=5 \times 10^{-5}$, and $k = N p$). The red triangles correspond to simulations of the full model, the blue circles to the {\it 3-D map with noise}, and the dashed line to the {\it 3-D map without noise}. The shaded grey region indicates parameter values for which the linear stability analysis predicts the fixed point $\bar \lambda = 1$ to be unstable. We observe that the {\it 3-D map with noise} captures the deviations from $\lambda = 1$ very well until these become large slightly past the onset of instability, i.e., approximately when $\sigma_\lambda \approx 0.025$. The {3-D map without noise}, neglecting fluctuations, fails to capture the small deviations from $\lambda = 1$ that occur before the onset of instability. To relate these findings with the distribution of avalanche sizes, we indicate  with an arrow the value of $C_1^*$ such that values $C_1 < C_1^*$ yield avalanche size distributions that have plausible power-law fits with exponent close to $-3/2$ (the same information that was used to color the curves in Fig.~\ref{fig:robustness_critical_state}). The map without noise thus predicts roughly the onset of instability and, correspondingly, of avalanche size distributions that are not power-law distributed.

%%%%%%%%%%%%%%%%%%%%%%%%%
%          3-DIMENSIONAL MAP: RESULT           %
%%%%%%%%%%%%%%%%%%%%%%%%%

\begin{comment}
\begin{figure}[b]
\centering{
\subfloat{{\label{fig:lambda_het_interaction}}\includegraphics[scale=0.35]{../used_figures/lambda_heterogeneous_sourcerates.eps}} \\
\subfloat{{\label{fig:R_het_interaction}}\includegraphics[scale=0.35]{../used_figures/R_heterogeneous_sourcerates.eps}}
}
\caption{Eigenvalue $\lambda$ as a function of time in the absence (red triangles) and presence (blue triangles) of diffusion between glial cells for the case of heterogeneous source rates. Diffusion between glial cells results in a robust critical state.}
\label{fig:heterogeneous_sourcerates}
\end{figure}
\end{comment}

%\begin{figure}
%\centering{
%\includegraphics[scale=0.35]{../used_figures/lambda_heterogeneous_sourcerates_v2.eps}
%}
%\caption{Eigenvalue $\lambda$ (a) as a function of time in the absence (red triangles) and presence (blue triangles) of diffusion between glial cells for the %case of heterogeneous source rates. Diffusion between glial cells results in a robust critical state.}
%\label{fig:heterogeneous_sourcerates}
%\end{figure}

%%%%%%%%%%%%%%%%%%%%%%%%%%%%%%%%%%%%
%    EXPERIMENT: HETEROGENEOUS GLIAL SOURCE RATES       %
%%%%%%%%%%%%%%%%%%%%%%%%%%%%%%%%%%%%

\begin{figure}
\centering{
\subfloat{{\label{fig:lambda_het_interaction}}\includegraphics[scale=0.35]{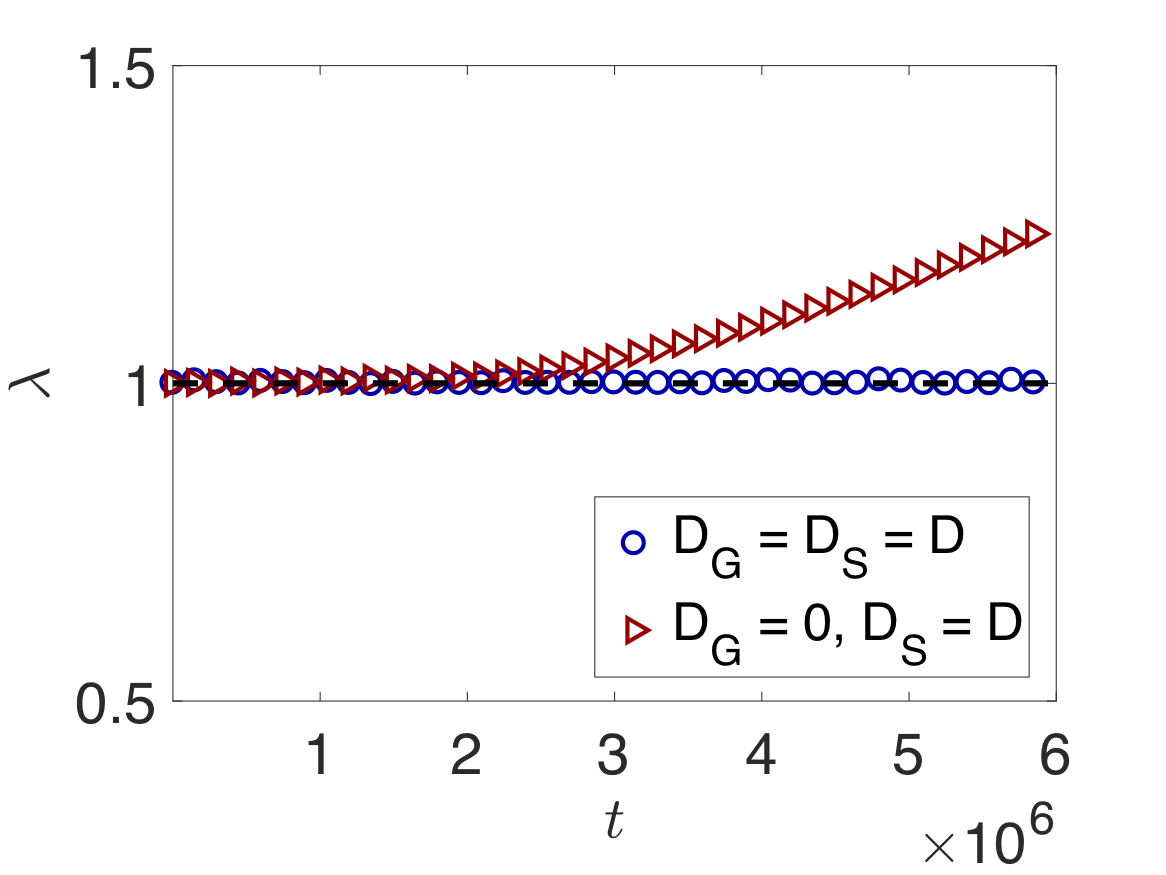}} \\
%\subfloat{{\label{fig:R_het_interaction}}\includegraphics[scale=0.35]{../used_figures/R_heterogeneous_sourcerates.eps}}
}
%\caption{Using heterogeneous interaction network, $G$, the critical state is stable only if resources are allowed to transport amongst the glial cells. Panel (a) shows that when there is diffusion amongst glia (blue circles), the largest eigenvalue of the neural network adjacency matrix, $\lambda \approx 1$. Thus, the critical state is stable. However, if resource transport amongst the glial cells is absent, the neural network becomes supercritical with $\lambda > 1$ (red triangles). The dashed line shows $\lambda=1$ for reference. Panel (b) shows that the total resource $\mathcal{R}$ reaches a steady state when diffusion is turned on (blue circles) and keeps increasing over time if diffusion is turned off (red triangles). This further highlights the importance of resource-transport dynamics on the stability of the critical state.}
\caption{Using heterogeneous source rates, $\lambda$ remains close to $1$ when there is diffusion amongst glia, i.e., $D_G = D$ (blue circles), and grows when resource transport amongst the glial cells is absent, i.e. $D_G = 0$. The dashed line shows $\lambda=1$ for reference.}
\label{fig:heterogeneous_sourcerates}
\end{figure}

So far, our results have been independent of resource transport in the glial network. The resource supply and consumption could be understood as a local homeostatic mechanism analogous to those discussed in Refs.~\cite{zierenberg2018homeostatic,kossio2018growing} and references therein.
%[e.g., $D_G$ does not appear in Eqs.~(\ref{eq:3DmapR}) and (\ref{eq:3Dmaplambda}), and numerical experiments show that our results so far hold if we set $D_G = 0$], and thus one might wonder if local supply and consumption of resources might be sufficient to lead the system towards a global critical state. 
However, in the next numerical experiment we show that, when there are heterogeneities (in the network structure, in the supply and consumption rates $C_1$ and $C_2$, etc), the diffusion of resources between glial cells can compensate for these effects and prevent destabilization of the critical state. We consider the particular case of heterogeneous source rates, where now each glial cell $i$ has its own $C_1^i$. As an example, we draw the $C_1^i$ from a Gaussian distribution with mean $C_1$ and standard deviation $2.6 \times 10^{-7}$, so that approximately $5\%$ of them do not satisfy the inequality $C_1^i < k C_2$. In the absence of resource transport, resource accumulates in these glial cells and the associated synapses, bringing the network to the supercritical state, $\lambda^t > 1$. However, when resource is allowed to diffuse, the critical state $\lambda^t \approx 1$ is maintained. This is shown in Fig.~\ref{fig:heterogeneous_sourcerates}, where we plot $\lambda^t$ as a function of time with $D_G = 0$ (red triangles) and $D_G = D_S > 0$ (blue circles). 
%Although not shown in the figure, the total resource $\mathcal{R} = \sum_i R_i + \sum_\eta R_\eta$ accumulates indefinitely when $D_G = 0$, and reaches a steady state when $D_G > 0$. 
We also note that in addition to stabilizing the critical state against parameter heterogeneities, network diffusion can stabilize the critical state in the presence of learning \cite{virkar:et:al:2016}.

\section{Discussion}

To summarize, we have found that resource-transport dynamics can stabilize the dynamics of excitable units so that they operate at a critical state characterized by experimentally-observable power-law distributed avalanche sizes. 
Using a reduced 3-dimensional map, we showed that for a large range of parameters the system self-organizes to a critical state that is characterized by power-law distributed avalanche sizes with an exponent value near the characteristic $-3/2$ exponent found in various experimental studies. We found that resource transport dynamics protects the system against the destabilizing effects of parameter heterogeneity. 
While our theoretical analysis is based on the assumption of a homogeneous network, it could potentially be generalized to account for heterogeneous or spatial network structure.
%and a spatially distributed transport network. 
%Finally, we emphasize that, 
Although we presented our model in the context of neuronal networks,
%, where the resource represents metabolites, 
our results could be applicable to other networks of excitable elements whose interaction efficacy depends on the availability of a shared resource.

\bibliographystyle{unsrt}
\bibliography{refs}

\end{document}